\documentclass[prb]{revtex4}
\usepackage[final]{graphicx}
\usepackage{dcolumn}

\begin{document}
\title{Quantum interference and Klein tunneling in graphene heterojunctions}
\author{Andrea F. Young and Philip Kim}
\affiliation{Department of Physics, Columbia University, New York, New York 10027, USA}
\date{\today}

\begin{abstract}
The observation of quantum conductance oscillations in mesoscopic systems has traditionally required the confinement of the carriers to a phase space of reduced dimensionality~\cite{vanwees_1989, Ji_Nature, liang_nature, lau_science}. While electron optics such as lensing~\cite{ Stormer_APL1990} and focusing~\cite{Beenakker_Focusing} have been demonstrated experimentally, building a collimated electron interferometer in two unconfined dimensions has remained a challenge due to the difficulty of creating electrostatic barriers that are sharp on the order of the electron wavelength~\cite{Washburn_PRB1988}. Here, we report the observation of conductance oscillations in extremely narrow graphene heterostructures where a resonant cavity is formed between two electrostatically created bipolar junctions. Analysis of the oscillations confirms that p-n junctions have a collimating effect on ballistically transmitted carriers.~\cite{katsnelson_natphys,falko_prb} The phase shift observed in the conductance fringes at low magnetic fields is a signature of the perfect transmission of carriers normally incident on the junctions~\cite{levitov_2008} and thus constitutes a direct experimental observation of ``Klein Tunneling.'' \cite{klein_1929}
\end{abstract}
\maketitle
Owing to the suppression of backscattering~\cite{ando_jpsj_1998_1} and its amenability to flexible lithographic manipulation, graphene provides an ideal medium to realize the quantum engineering of electron wave functions. The gapless spectrum in graphene~\cite{novoselov_science} allows the creation of adjacent regions of positive and negative doping, offering an opportunity to study the peculiar carrier dynamics of the chiral graphene quasiparticles~\cite{klein_1929,katsnelson_natphys,falko_prb} and a flexible platform for the realization of a variety of unconventional electronic devices.~\cite{cheianov_science,park_natphys,park_nl,garcia_prl} Previous experiments on graphene p-n junctions~\cite{lemme_ieee_2007,huard_prl,williams_science,ozyilmaz_prl,liu_apl,gorbachev_nanolett,stander_arxiv_2008,meric_naturenano} were limited in scope by the diffusive nature of the transport beneath the local electrostatic gates; we overcome such limitations by fabricating extremely narrow ($\sim$20~nm) local gates strongly capacitively coupled to the graphene channel (Fig.~1a-b). Electrostatics simulations based on finite element analysis (see online supplementary material) show that the carrier densities in the locally gated region (LGR) and the `graphene leads' (GL)---$n_2$ and $n_1$, respectively---can be controlled independently by applying bias voltages to the top gate ($V_{TG}$) and the back gate ($V_{BG}$). The width of the LGR, $L$, is defined as the distance between the two zero density points. As in previous studies~\cite{ ozyilmaz_prl}, the conductance map as a function of $V_{TG}$ and $V_{BG}$ (Fig.~1c) can be partitioned into quadrants corresponding to the different signs of $n_1$ and $n_2$, with a lowered conductance observed when $n_1n_2<0$. The mean free path in the bulk of the sample, $l_m\gtrsim$100~nm, was extracted from the relation $\sigma=\frac{2e^2}{h}k_Fl_m$ between the conductivity and Fermi momentum, $k_F$.  Since $L\lesssim100$~nm within the experimentally accessible density regime, we expect a significant portion of the transport to be ballistic in the LGR.

In the bipolar regime, the diffusive resistance of the LGR is negligible in comparison with the highly resistive p-n junctions; as a result, the conductance does not increase with increasing magnitude of the charge density in the LGR.\cite{gorbachev_nanolett} We note that the magnitude of this conductance step is only $\sim60\%$ as large as expected for a fully ballistic heterojunction even after taking into account the enhancement of the junction transparency due to nonlinear screening;\cite{fogler_prl} this suggests that there is still a large diffusive component to the transport through the heterojunction. Nevertheless, each trace exhibits an oscillating conductance as a function of $V_{TG}$ when the carriers in the LGR and GL have opposite sign.

The regular structure of these oscillations is apparent when the numerical derivative of the measured conductance is plotted as a function of $n_1$ and $n_2$ (Fig.~2a). While there is a weak dependence of the oscillation phase on $n_1$ reflecting the influence of the back gate on the heterojunction potential profile, the oscillations are primarily a function of $n_2$, confirming their origin in cavity resonances in the LGR. The oscillations, which arise due to interference between electron waves in the LGR, are not periodic in any variables due to the strong dependence of the LGR width, $L$, and junction electric field, $E$, on the device electrostatics. Still, the conductance maxima are separated in density by roughly $\Delta~n_2\sim 1\times 10^{12}$~cm$^{-2}$, in agreement with a naive estimate $\Delta~n_2\sim \frac{4\sqrt{\pi n_2}}{L}$ for the resonant densities in a cavity of width $L\sim100$~nm. The application of an external magnetic field shifts the phase of the oscillations, with individual oscillation extrema moving towards higher density $|n_2|$ and the transmission resonances appearing to be adiabatically connected to the high field Shubnikov-de Haas oscillations (Fig.2d).

Graphene heterojunctions offer the opportunity to study an old problem in relativistic quantum mechanics: the tunneling of relativistic electrons through a potential barrier.~\cite{klein_1929,katsnelson_natphys} In the context of the graphene p-n junction, this ``Klein tunneling'' manifests as the combination of the absence of backscattering with momentum conservation parallel to a straight p-n interface: normally incident particles, bound to conserve their transverse momentum, $k_y=0$, and forbidden from scattering directly backwards, are predicted to tunnel through such symmetric potential barriers with unit probability. In contrast, particles obliquely incident on a barrier which is smooth on the atomic lattice scale encounter classically forbidden regions where the real part of the perpendicular momentum vanishes.  These regions, which form about the center of individual p-n junctions, transmit obliquely incident carriers only though quantum tunneling, leading to an exponential collimation of ballistic carriers passing through graphene pn junctions,~\cite{falko_prb}

\begin{equation}
\left|T(k_y)\right|^2=e^{-\pi\hbar v_F k_y^2/(eE)}
\label{fal}
\end{equation}
where $v_F$ is the Fermi velocity of graphene.

Considerable experimental effort has been expended trying to verify eq. (\ref{fal}) by matching bulk resistance measurements across an p-n junctions with their expected values~\cite{gorbachev_nanolett,stander_arxiv_2008,huard_prl}. Such an approach can, at best, provide indirect evidence for the theoretically predicted features of chiral tunneling---collimation and perfect transmission at normal incidence.  In particular, there is no way to distinguish perfect from near perfect transmission from a bulk resistance measurement, which is sensitive only to the total transparency of the pn junction.  The quantum interference experiments presented in this letter allow a measurement not only of the magnitude but also the phase of the transmission and reflection coefficients. Interestingly, whereas the bulk of conduction in a fully ballistic graphene p-n junction is expected to be dominated by normally incident carriers, the absence of backscattering precludes the contribution of such trajectories to the Fabry-Perot resonances due to perfect normal transmission at both interfaces.  Rather, the oscillatory conductance receives its largest contributions from particles incident at angles where neither the transmission probability, $|T^2|$, nor the reflection probability, $|R|^2=1-|T|^2$, are too large (see, e.g.,  marker 1 in Fig.~2c). Only transmission near such angles contributes to the oscillatory conductance, ensuring the survival of the oscillations despite the incident angle averaging and allowing the determination of the width of the angle of acceptance of an individual collimating p-n junction.

In a ballistic heterojunction, the application of a magnetic field bends the carrier trajectories, resulting in an addition of an Aharonov-Bohm phase to the interference and a modification of the angle of incidence at each pn junction. As was pointed out recently~\cite{levitov_2008}, such cyclotron bending provides a direct experimental signature of reflectionless tunneling, which manifests as a phase shift in the transmission resonances of a ballistic, phase coherent, graphene heterojunction at finite magnetic field. These resonances are described by the etalon-like ray tracing diagrams shown in Fig 2b. The Landauer formula for the oscillating part of the conductance is then

\begin{equation}
G_{osc}=e^{-2L/l_{LGR}}\frac{4e^2}{h}\sum_{k_y} 2|T_+|^2|T_-|^2|R_+||R_-|\cos\left(\theta_{WKB}+\Delta\theta_{rf}\right), \label{Gosc}
\end{equation}
where $T_\pm$ and $R_\pm$ are the transmission and reflection amplitudes for the classically forbidden regions centered at $x=\pm L/2$, $\theta_{WKB}$ is the semiclassical phase difference accumulated between the junctions by interfering trajectories, $\Delta\theta_{rf}$ is Klein backreflection phase of the two interfaces, and $\l_{LGR}$ is the mean free path in the locally gated region, a fitting parameter which controls the amplitude of the oscillations (see Supplementary Information).

At zero magnetic field, particles are incident at the same angle on both junctions, and the Landauer sum in eq. (\ref{Gosc}) is dominated by modes which are neither normal nor highly oblique, as described above.  As the magnetic field increases, cyclotron bending favors the contribution of modes with $k_y=0$, which are incident on the junctions at angles with the same magnitude but opposite sign (see marker 2 and 3 in Fig.2c).  In the case of perfect transmission at zero incident angle, the reflection amplitude changes sign as the sign of the incident angle changes,~\cite{levitov_2008} causing a $\pi$ shift in the phase of the reflection amplitudes.  Equivalently, this effect can be cast in terms of the Berry phase: the closed momentum space trajectories of the modes dominating the sum at low field and high $k_y$ do not enclose the origin, while those at intermediate magnetic fields and $k_y\sim0$ do (Fig. 2b). Due to the Dirac spectrum and its attendant chiral symmetry, there is a topological singularity at the degeneracy point of the band structure, $k_x=k_y=0$, which adds a non-trivial Berry phase of $\pi$ to trajectories surrounding the origin.  As a consequence, the quantization condition leading to transmission resonances is different for such trajectories, leading to a phase shift in the observed conductance oscillations (i.e., a $\pi$ jump in $\Delta\theta_{rf}$) as the phase shifted trajectories begin to dominate the Landauer sum in Eq.~(\ref{Gosc}).~\cite{zhang_nature,novoselov_nature} For the electrostatics of the devices presented in this letter, the magnetic field at which this phase shift is expected to occur is in the range $B^*\sim$250--500 mT (see Supplementary Information), in agreement with experimental data (see Fig. 3a).  As the magnetic field increases further, the ballistic theory predicts the disappearance of the Fabry-Perot conductance oscillations as the cyclotron radius, $R_c$, shrinks below the distance between p-n junctions, $R_c\lesssim L$, or B$\sim$ 2~T for our devices.  We attribute the apparent continuation of the oscillations to high magnetic field to the onset of disorder mediated Shubnikov-de Haas type oscillations within the LGR.

In order to analyze the quantum interference contribution to the ballistic transport, we extract the oscillating part of the measured conductance by first antisymmetrizing the heterojunction resistance~\cite{huard_prl} with respect to the density at the center of the LGR, $G_{odd}^{-1}(|n_2|)=G^{-1}(n_2)-G^{-1}(-n_2)$, and then subtracting a background conductance obtained by averaging over several oscillation periods in $n_2$, $G_{osc}=G_{odd}-\overline{G_{odd}}$.  The resulting fringe pattern shows a marked phase shift at low magnetic field in accordance with the presence of the Klein backscattering phase, with two different regions---of unshifted and shifted oscillations---separated by the magnetic field $B^*$ (see Fig 3a). To perform a quantitative comparison between the measured $G_{osc}$ and eq. (\ref{Gosc}), we first determine the potential profile in the heterojunction devices from numerical electrostatics simulations, which information is then input into (\ref{Gosc}) to generate a fringe pattern for comparison with experimental data.  We choose the free fitting parameter $l_{LGR}=67$~nm for this comparison to best fit the oscillation amplitudes.  Considering possible degradation of the graphene in and around the LGR during the fabrication of the local gates,~\cite{Sunmin} this value is consistent with the estimate for the bulk mean free path.  The resulting theoretical calculation exhibits excellent quantitative agreement with the experimental result at both zero and finite magnetic field (Fig. 3a-c) both in the magnitude and period of the oscillations.  We emphasize that the value of $L$---which largely determines both the phase and amplitude of the oscillations---varies by almost by a factor of three over the accessible density range, yet Eq. (\ref{Gosc}) faithfully describes the observed experimental conductance modulations in $n_2$ as well as in $B$. Such remarkable agreement confirms that the observed oscillatory conductance, which is controlled both by the applied gate voltage and the magnetic field, results from quantum interference phenomena in the graphene heterojunction.  Moreover, the oscillations exhibit a phase shift at  $B^*\sim$~0.3~T which is the hallmark of perfect transmission at normal incidence, thus providing direct experimental evidence for the ``Klein tunneling'' of relativistic fermions through a potential barrier.

Finally, we turn our attention to the temperature dependence of the quantum coherence effects described in the text, which we observe at temperatures as high as 60~K (Fig.~3d).  An elementary energy scale analysis suggests that the phase coherence phenomena should be visible at temperatures of order $\frac{\hbar v_F}{L}\sim 100$~K, when thermal fluctuations become comparable to the phase difference between interfering paths. In addition, the oscillation amplitude is sensitive to the carrier mean free path, and we attribute the steady waning of the oscillations with temperature to a combination of thermal fluctuations and further diminution of the mean free path by thermally activated scattering. The mean free path in clean graphene samples can be as large as $\sim$1~$\mu$m,~\cite{bolotin_ssc} and a reduction of the width of the heterostructure $L$ by an order of magnitude is well within the reach of modern fabrication techniques; consequently, technological improvements in the fabrication of graphene heterojunctions should lead to the observation and control of quantum coherent phenomena at much high temperatures, a crucial requirement for realistic, room temperature quantum device applications.
\subsection*{Methods}
Graphene sheets were prepared by mechanical
exfoliation\cite{novoselov_science} on Si wafers covered in 290~nm
thermally grown SiO$_2$. Ti/Au contacts 5~nm/35~nm thick were
deposited using standard electron beam lithography, and local
gates subsequently applied using a thin ($\sim10$ nm)
layer of hydrogen silsesquioxane (HSQ) as an adhesion
layer~\cite{ozyilmaz_prl} for low-temperature atomic layer deposition of 20 nm
of HfO$_2$, a high-$k$ dielectric ($\epsilon\sim 12$) (see Fig.~1b).
Palladium top gates not exceeding 20 nm in width were deposited in
order to ensure that a sizeable fraction of conduction electrons
remained ballistic through the LGR.  Leakage current was measured
to be $\leq 100$~pA up to $V_{TG}=\pm15$~V. All data except Fig.~
3d was taken from the device depicted in Fig 1a, which had a
measured mobility $\sim$~5,000 cm$^2$/V sec.
Fig.~3d was taken from a similar device in a four terminal Hall bar geometry;
additional data from this device is shown in the supplementary materials.
Several other similar devices were also measured, showing qualitatively similar behavior.  The conductance of the graphene
devices was measured in a liquid helium flow cryostat at 4.2-
100~K using a standard lock-in technique with a current bias of
.1-1 $\mu$A$_{rms}$ at 17.7 Hz. Unless otherwise specified, all
measurements were done at 4.2~K. The ratio $C_{TG}/C_{BG}\approx12.8$
was determined from the slope of the Dirac ridge with respect to
the applied voltages, and similar values were obtained from the
analysis of the period of the Shubnikov-de Haas oscillations in magnetic field,
which also served to confirm the single layer character of the
devices.  Finite element electrostatics simulations were carried
out for the measured device geometries described above with the
thickness and dielectric constant of the HSQ adjusted such that
the simulations matched the observed values of $C_{TG}/C_{BG}$.
The shape of the potential and the strength of the electric field $E$ used in fitting the experimental data
were constrained to lie within the confidence interval of the
simulations, which in turn were largely determined by uncertainty
in the device geometry.

\begin{acknowledgments}
The authors would like to thank I.L. Aleiner, K.I. Bolotin, M.Y.
Han, E.A. Henriksen, L.S. Levitov, and H.L. Stormer for discussions, and I. Meric
and M.Y. Han for help with sample preparation.  This work is
supported by the ONR (No. N000150610138), FENA, NRI, and NSEC (No.
CHE-0117752), and NYSTAR. Sample preparation was supported by DOE
(DE-FG02-05ER46215).
\end{acknowledgments}

\newpage\begin{figure}[h]\includegraphics[width=120mm]{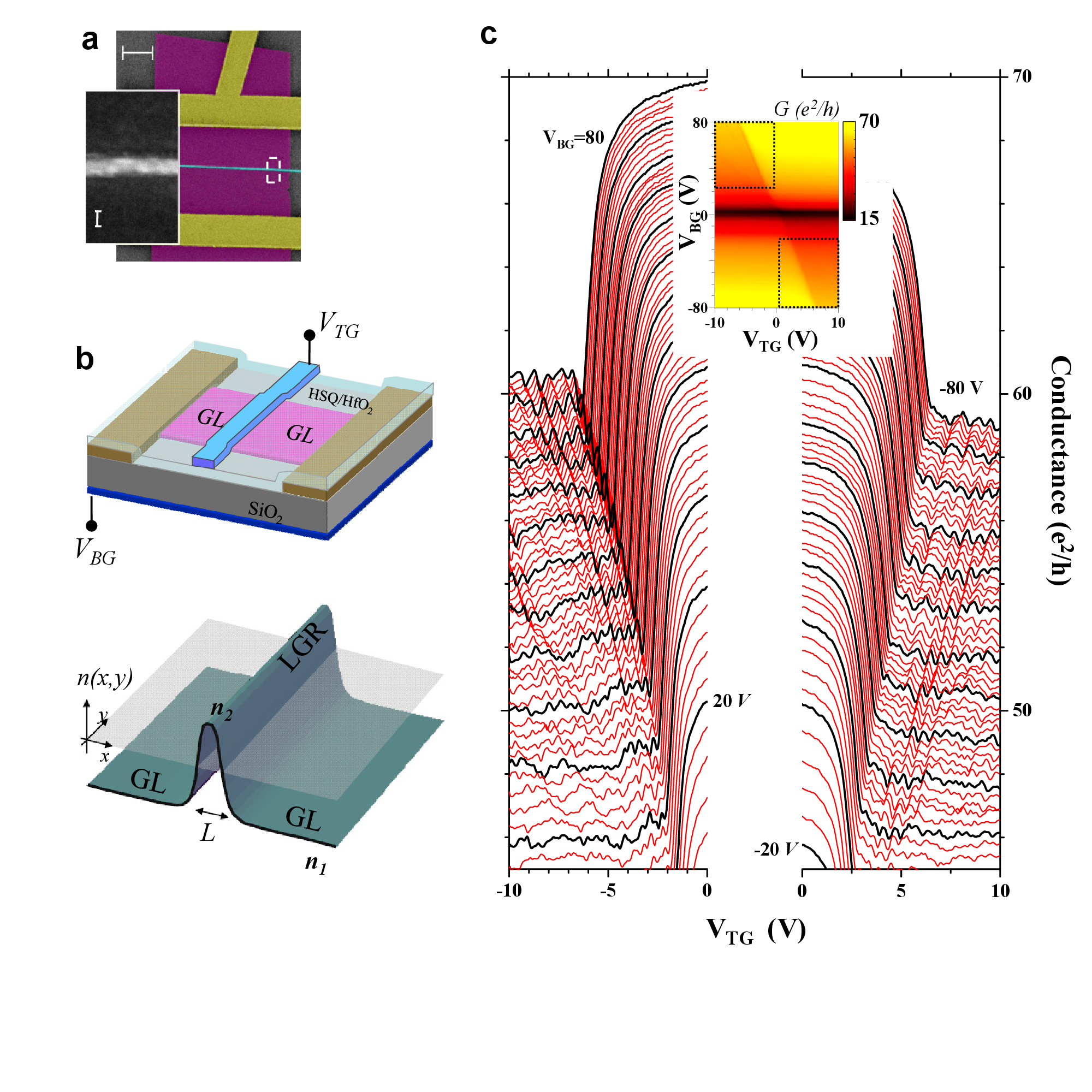}\caption{Graphene heterojunction device schematics and conductance measurements. {\bf a} False color scanning electron microscope image of a typical graphene heterojunction device. Electrodes, graphene, and top gates are represented by yellow, purple and cyan, respectively. The scale bar is 2~$\mu$m. Inset: high magnification view of top gate. The scale bar is 20~nm. {\bf b} Schematic diagram of the device geometry. The electrostatic potential created by the applied gate voltages, $V_{BG}$ and $V_{TG}$, can create a graphene heterojunction of width $L$ bounded by two p-n junctions. {\bf c} The inset shows the conductance as a function of $V_{TG}$ and $V_{BG}$. The main panels show cuts through this color map in the regions indicated by the dotted lines in the inset, showing the conductance as a function of $V_{TG}$ at fixed $V_{BG}$. Traces are separated by step in $V_{BG}$ of 1 V, starting from $\pm 80$ with traces taken at integer multiples of 5~V in black for emphasis.} \end{figure}

\newpage\begin{figure}[h]\includegraphics[width=120mm]{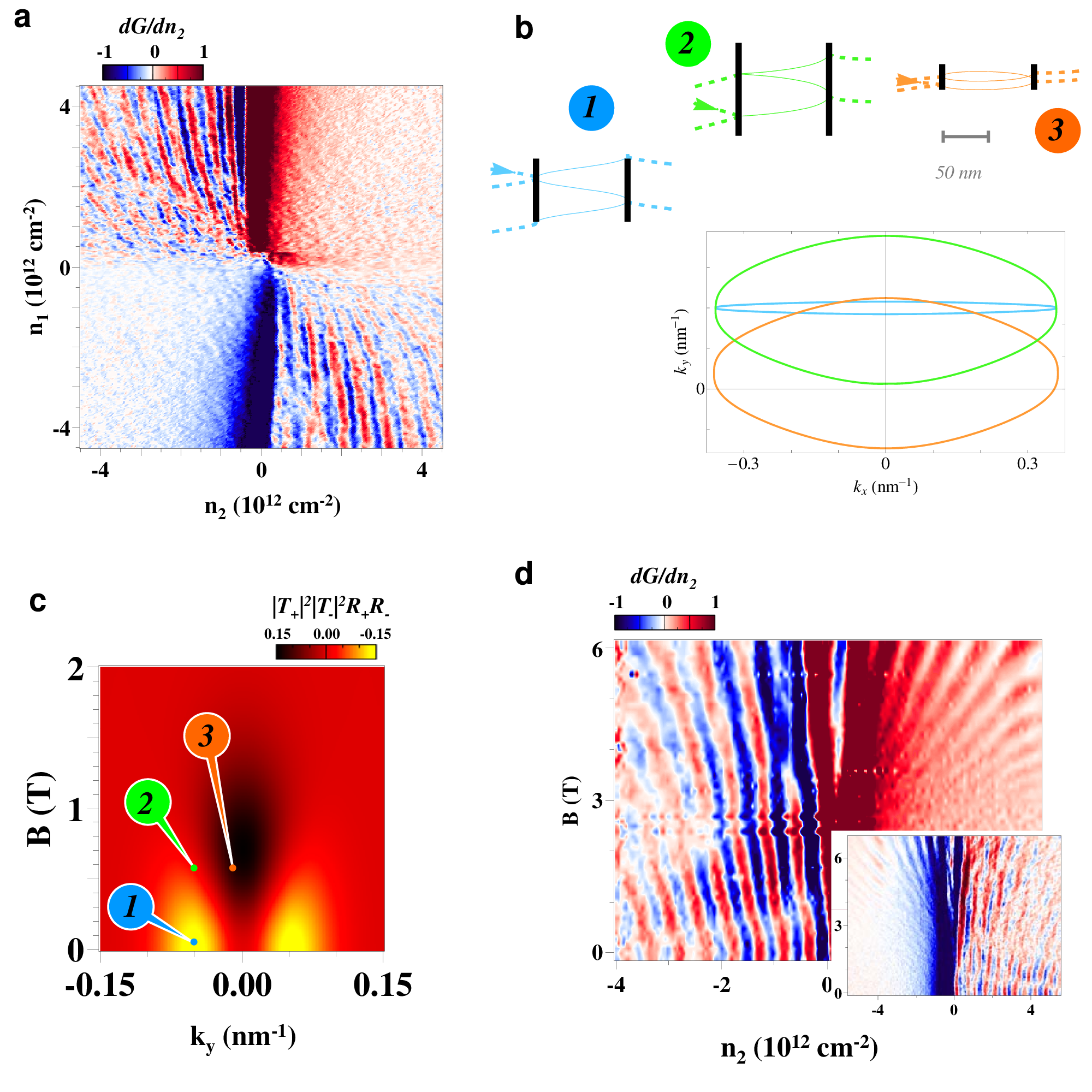} \caption{{\bf a}  $dG/dn_{2}$ as a function of $n_1$ and $n_2$.  Scale bar is in arbitrary units. {\bf b} Schematic of trajectories contributing to quantum oscillations in real and momentum space.  The dominant modes at low magnetic field (marker 1) give way, with increasing $B$, to phase shifted modes with negative reflection amplitude due to the inclusion of the non-trivial Berry phase (marker 3), near $k_y=0$.  The original finite $k_y$ modes are not yet phase shifted at this field (marker 2), but due to collimation, they no longer contribute to the oscillatory conductance.
{\bf c} The prefactor in the Landauer sum, $|T_+|^2|T_-|^2R_+R_-=|T_+|^2|T_-|^2|R_+||R_-|e^{i\Delta\theta_{rf}}$, as a function of $B$ and $k_y$, plotted for the experimental parameters at V$_{BG}$=50V for $n_2=3.5 \times 10^{12}$~cm$^{-2}$. The markers correspond to the trajectories shown in b.  Regions of negative sign correspond to trajectories containing the Klein backscattering phase shift.  {\bf d} Main panel:  Magnetic field dependence of $dG/dn_{2}$ at $V_{BG}=50$~V.  Inset: Similar data taken at $V_{BG}=-50$~V.  The magnetic phase is proportional to the sign of the carriers; as a result, the oscillation extrema precess in opposite directions for opposite signs of carriers in the LGR.} \end{figure}

\newpage\begin{figure}[h]\includegraphics[width=100mm]{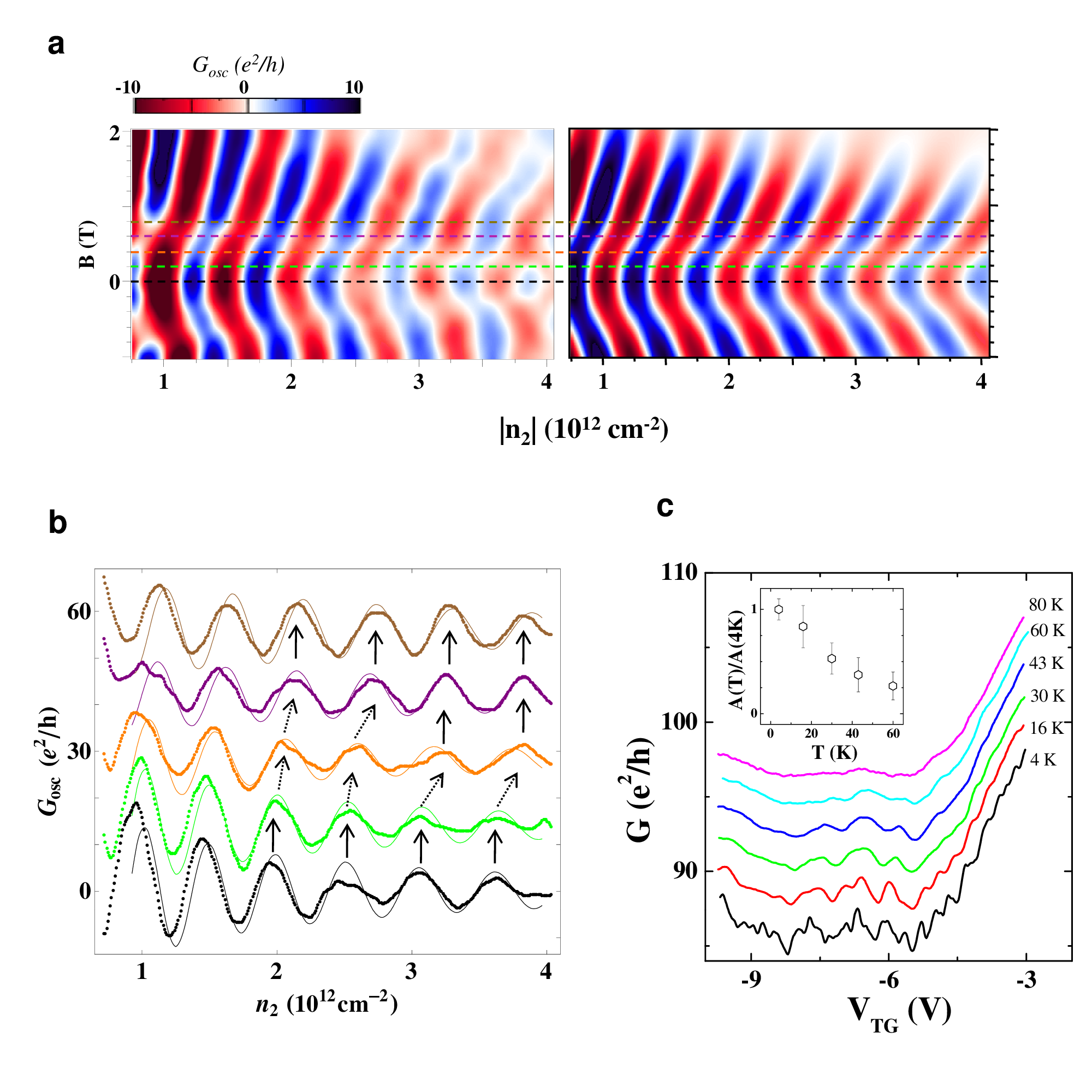}\caption{Comparison of experimental data with the theoreticail model, and temperature dependence.  {\bf a} Magnetic field and density dependence of the oscillating part of the conductance at $V_{BG}=50$~V.  $G_{osc}$ as extracted from the experimental data (left panel) shows good agreement with a theoretical model accounting for nonlinear screening\cite{fogler_prl} (right panel) over a wide range of densities and magnetic fields. {\bf b} Cuts taken at $B=0$, 200, 400, 600, and 800 mT, corresponding to the colored dashed lines in a; the dots represent data, the smooth lines the result of the simulations. The sudden phase shift that signals the presence of perfect transmission is indicated by dashed arrows. Curves are offset for clarity.  {\bf c} Temperature dependence of the oscillation amplitude in a similar device to that presented in the rest of the text. Main panel: The oscillations (different curves are offset for clarity) weaken with rising temperature, and are not observed above 80~K. At 4~K, the conductance modulations contain both the ballistic oscillations as well as aperiodic modulations due to universal conductance fluctuations which quickly disappear with increasing temperature. Inset:  Averaged amplitude of several oscillations, normalized by the amplitude at $T=4$~K.} \end{figure}

\newpage
\section*{Supplementary Information}
\subsection*{Details of theoretical model}
The Landuaer formula for the oscillating part of the conductance is obtained from the ray tracing scheme shown in Figure 2a of the main text.  The transmission amplitude through the entire junction is taken to be a product of the transmission amplitudes at the two interfaces with a phase factor corresponding to the semiclassical phase accumulated between the junctions.  This semiclassical phase difference for neighboring trajectories is
\begin{equation}
\theta_{WKB}=\Re \int_{-L/2}^{L/2}\sqrt{\pi |n(x)|-\left(k_y-\frac{e}{\hbar}Bx\right)^2},\label{wkb}
\end{equation}
where we take the real part to account for the fact that in general the classical turning points shift from their values of $\pm L/2$ defined for $k_y=0$, $B=0$.  In addition, there is a nonanalytic part of the phase associated with the the vanishing of the reflection coefficients at opposite interfaces~\cite{levitov_2008} which can be nontrivial at finite magnetic field,
\begin{equation}
\Delta\theta_{rf}=\pi\left(H\left(-k_y+\frac{eBL}{2\hbar}\right)-H\left(-k_y-\frac{eBL}{2\hbar}\right)\right),
\end{equation}
where $H(x)$ is the step function ($H(x)=1$ for $x>0$, $H(x)=0$ otherwise).  This phase jump is equivalent to a sign change in the reflection coefficient as the incidence angle crosses zero, and implies that the transmission probability at normal incidence is unity.

At a p-n junction, assumed to be smooth on the lattice scale, the transmission amplitude is exponentially peaked about normal incidence.~\cite{falko_prb} The principal effect of a weak magnetic field on the transmission through a single p-n junction is to modify the incident angle at the two junctions due to cyclotron bending of the trajectories.  Choosing the Landau gauge where $\vec A= B x \hat y$ with $k_y$ the conserved transverse momentum in the center of the junction ($x=0$), the transmission and reflection amplitudes at the junctions located at $x=\pm L/2$ are~\cite{levitov_2008}

\begin{equation}
T_\pm =\exp\left(\frac{\pi\hbar v_F}{2eE} \left(k_y\pm \frac{eBL}{2\hbar}\right)^2\right)\qquad R_\pm=\exp\left(i\pi H\left(-k_y\mp eBL/(2\hbar)\right)\right)\sqrt{1-|T_\pm|^2}
\label{trans}
\end{equation}

Defining the total phase $\theta=\theta_{WKB}+\Delta \theta$, and taking into account the damping, due to scattering, of the particle propagators between the junctions, we can write the Landauer conductance of the heterojunction via the canonical Fabry-Perot formula as
\begin{equation}G=\frac{4e^2}{h}\sum_{k_y} \left|\frac{T_+T_- e^{-L/(2l_{LGR})}}{1-|R_+||R_-|e^{i\theta}e^{-L/l_{LGR}}}\right|^2 \label{Gtot}
\end{equation}

It is difficult to separate diffusive from ballistic effects in the bulk conductance.  However, the contribution of diffusive effects to quantum interference effects are strongly suppressed at $B=0$, rendering useful the definition $G_{osc}\equiv G-\overline{G}$, where $\overline{G}$ denotes the conductance averaged over the accumulated phase.  Multiple reflections are suppressed both by the finite mean free path as well as the collimating nature of the junctions; as $|T_\pm|,|R_\pm|\leq1$, higher order products of transmission and reflection coefficients are necessarily decreasing.  Utilizing this fact, we can expand the denominator in eq. (\ref{Gtot}) and subtract all nonoscillating terms to get the leading contribution to the quantum interference,

\begin{equation}
G_{osc}= \frac{8e^2}{h}\sum_{k_y} |T_+|^2|T_-|^2|R_+||R_-|\cos\left(\theta\right)e^{-2L/l_{LGR}}. \label{Goscsupp}
\end{equation}

This formula does not take into account inhomogeneities in the applied local gate potential due to the uneven width of the top gate or thickness and crystallinity of the dielectric.  Such disorder can lead to significant damping of the oscillations even in a completely ballistic sample~\cite{levitov_2008}, and makes the estimate for the mean free path derived from matching the observed amplitude of the oscillations a lower bound.  Nevertheless, that the mean free path in the LGR is shorter than that in the GL is consistent with recent experiments (unpublished) which find a strong enhancement of the Raman D-band in graphene after it is covered in HSQ and irradiated with electrons at energies and doses comparable to those used during top gate fabrication.~\cite{Sunmin}

\subsection*{Electrostatics simulations}
The theoretical model described in the previous section takes as an input the potential profile across the heterojunction. In order to perform a quantitative comparison between experiment and theory, we first determine these parameters by numerical simulation of our device electrostatics using Comsol Multiphysics, a commercial finite element simulation software package.  Scanning electron microscopy images of the device chosen for the quantitative comparison show the top gates to be $\sim20$~nm wide, while the capacitive coupling of the top gate is found to be $C_{TG}\approx 12.8 C_{BG}$.  Graphene is treated as a perfect conductor covered by 10-20~nm of hydrogen silsesquioxane (HSQ), with dielectric constant $\epsilon\sim2-5$, and 20~nm of HfO$_2$, $\epsilon\sim10-15$.  We choose the dielectric constant of the HSQ in each simulation to ensure a match between the calculated and measured capacitive coupling, determined as the relation between the applied top gate voltage and the maximal density reached in the LGR. The dielectric constant of atomic layer deposition (ALD) grown HfO$_2$ layer was consistent with a separate capacitance measurement on a similarly prepared thin film. For this range of sample parameters, we find the density profile in the heterojunction to be well approximated by
\begin{equation}
n(x)=\frac{C_{TG} V_{TG}}{1+\left|x/w\right|^{2.5}} + C_{BG} V_{BG},
\label{n}\end{equation}
where $w\sim$~45-47~nm is the effective width of the potential and the gate potentials $V_{TG}$ and $V_{BG}$ are coupled to the charge density through the capacitances $C_{TG}\approx$~1490~aF$\mu$m$^{2-}$ and $C_{BG}\approx$116~aF$\mu$m$^{-2}$. Here the exponent of $x/w$ is chosen empirically, and produces a potential profile that deviates by less than 10\% from that produced by the electrostatics simulations.

Fig.~4a shows density profiles in the graphene heterojunction which produce, via the Landuaer formula described in the previous section, conductance fringes matching those observed in experiments done at $V_{BG}=50$~V.  From these density profiles, the potential profile, distance between classical turning points, and density gradient at the center of the p-n junctions is determined.  We note that the distance between pn junctions, which determines the boundary of integration in equation (\ref{wkb}), changes by a factor of two over the experimentally accessed density range, making the excellent quantitative fits obtained for the oscillation phase particularly convincing.

\newpage \begin{figure}[h] \centering \includegraphics[width=120mm]{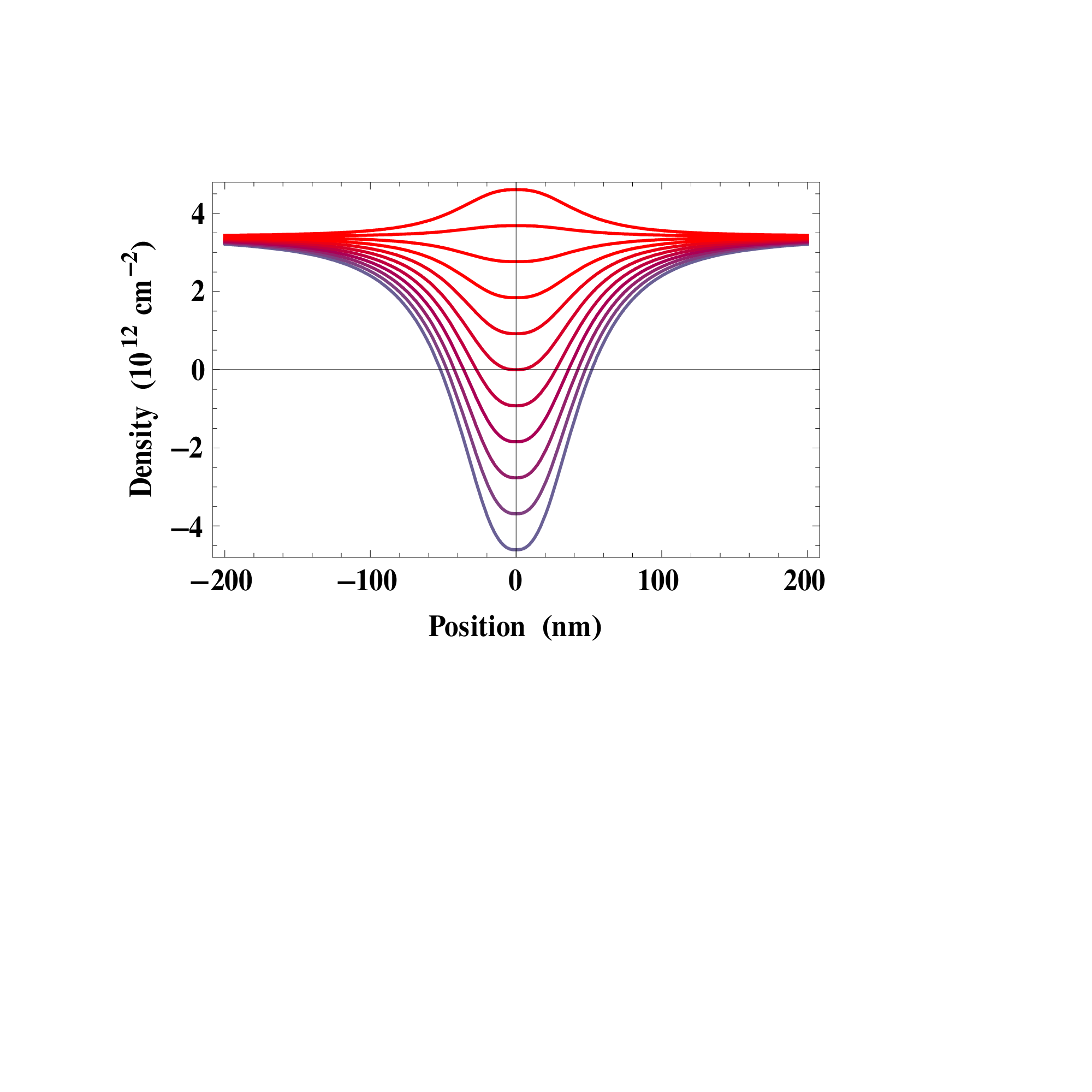} \caption{Density profiles for $V_{BG}=50$~V over a range of top gate voltages (from eq. {\ref{n}}).  Top gate voltage in 1~V increments.  The width of the central region ($L$) depends strongly on applied voltage. }\label{figsel}\end{figure}

\newpage
\subsection*{Quantitative estimate of the electric field at the p-n interfaces}
In this section, we compare the measured data (Fig. 5a) with numerical simulations to extract the strength of the electric field in an individual pn junction.  It follows from robust semiclassical arguments~\cite{falko_prb} that, in the experimentally realized situation of p-n junctions smooth on the scale of the lattice constant, the collimation at an individual junction should be a Gaussian function of $k_y$ (see eq .\ref{trans}).  The important parameter is the electric field, $E$, in the junctions, which is given, after taking into account the absence of linear screening in graphene near the charge neutrality point, by~\cite{fogler_prl}
\begin{equation}
eE=2.1 \hbar v_F n'^{2/3}. \label{fogler}
\end{equation}
where $n'$ is the density gradient across the junction.  As is evident from the simulations (Fig. 5b), the nonlinear screening correction to the electric field gives a better fit to the experimental data than either the non-exponential collimation produced by atomically sharp barriers---which appears to contain higher order resonances---or the weaker field that results from neglecting nonlinear screening near the Dirac point.

To make this comparison more quantitative, we perform several simulations in which the nonlinear screening result, (\ref{fogler}), is scaled by some prefactor, $\eta$.  For a rough comparison, the linear screening corresponds to $\eta\lesssim.5$, while the step potential corresponds to $\eta \gg20$.  As explained in the main text, the magnetic field dependence contains an abrupt phase shift at finite magnetic field as the finite $k_y$ modes cease to contribute to the oscillations and $k_y=0$ modes---which carry with them the additional Berry phase of $\pi$---become the dominant contribution to the oscillatory conductance.  With increasing magnetic field, a fully ballistic model predicts the gradual ebbing of these phase shifted oscillations as the cyclotron radius becomes comparable to the junction size.  The field at which the $\pi$-phase shift manifests is tied to the degree of collimation of the transmission at each p-n interface. Because this phase shift is rather abrupt, we can define the transition magnetic field, $B^*$, as the field at which the values of the the oscillation prefactor $|T_+|^2|T_-|^2|R_+||R_-|$ for zero and finite $k_y$ modes become comparable, giving $B^*\propto\sqrt{\frac{\hbar e E}{e^2 v_F L^2}}$. Since $B^*$ depends strongly on the junction electric field, it allows us to extract this field from the experimentally observed oscillation phase shift.  Defining $B^*$ as the field at which $G_{osc}(n_2, B=B^*)=0$ for fixed density $n_2$ such that $G_{osc}(n_2, B=0)$ is an extremum, we can estimate $\eta=.9\pm.3$.

In accordance with the ballistic theory, the oscillations peak at zero magnetic field, and then have a second maximum after the phase shift at finite magnetic field.  The relative height of these two maxima can be used to estimate the electric field $E$. Higher collimation suppresses the contribution of the modes near $k_y=0$ at finite $B$, since this feature is entirely generated by modes not normally incident at either interface. Higher collimation thus corresponds to an effectively more one-dimensional channel for interference effects, leading to the more effective destruction of the oscillations by the Lorentz force, which serves to push the particles out of the narrower acceptance angles at each junction.  By taking the average value of $G_{B}^{Max}/ G_{0}^{Max}$ and comparing it with the simulations for a variety of values of $\eta$, we can extract $\eta \approx$1.6$\pm$.3. We note that the apparent discrepancy between this and the method previously outlined is likely due to the contribution of momentum non-conserving processes to the oscillations at finite $B$, leading to an overestimate of $\eta$.  While neither of the methods above is free of systematic errors, they confirm the importance of nonlinear screening in determining transport through graphene p-n junctions.

\newpage \begin{figure}[h] \centering \includegraphics[width=120mm]{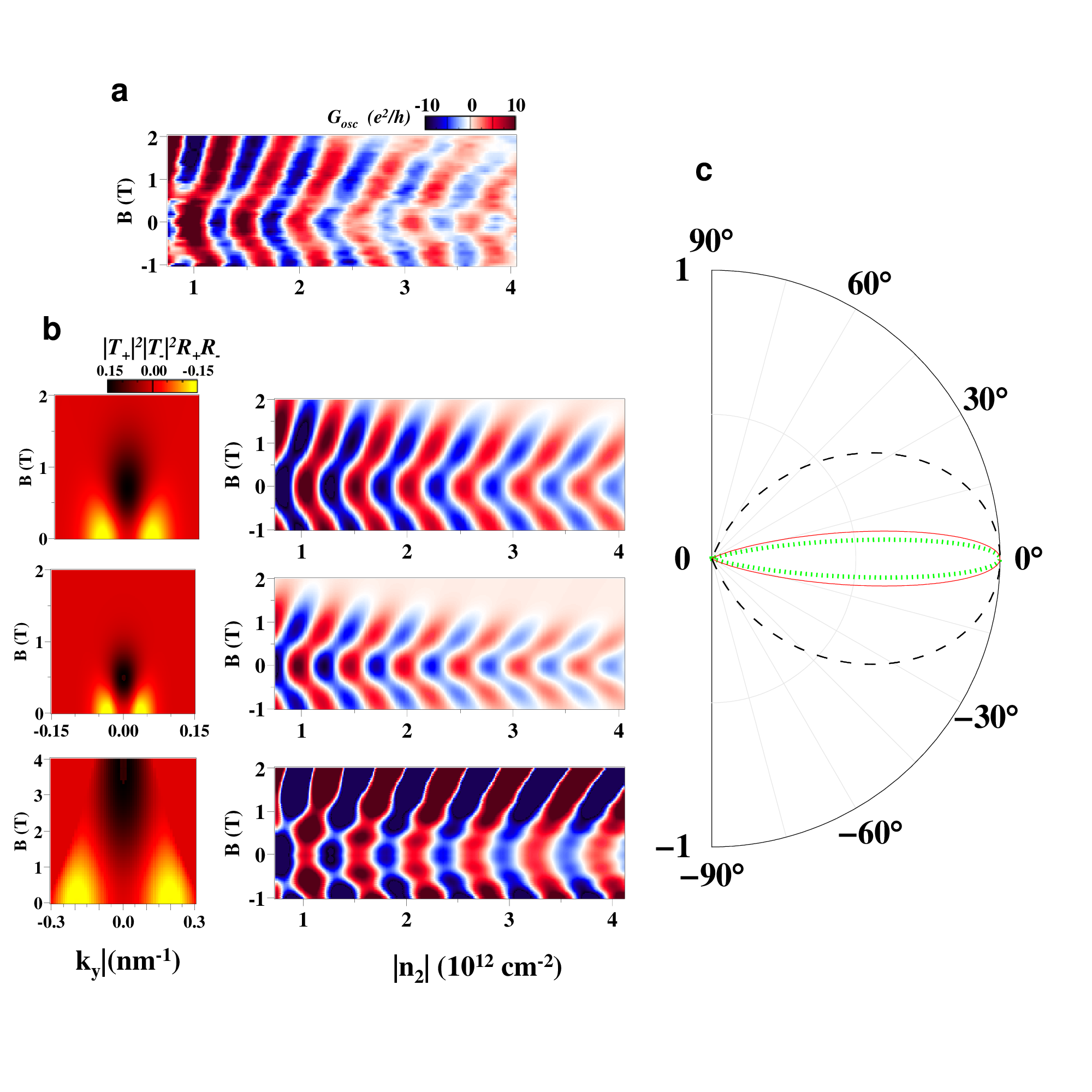} \caption{{\bf a} Top panel: oscillatory conductance as a function of $n_2$ and magnetic field at V$_{BG}$=50V.  {\bf b} The oscillation prefactor $|T_+|^2|T_-|^2R_+R_-$ (column 1) and resulting oscillatory conductance (column 2) as a function of magnetic field for a variety of collimation models.  The best fit to the data is achieved by accounting for the nonlinear screening (top panel); the simulations resulting from naively linearizing the potential between the extrema (middle panel) and considering the algebraic collimation resulting from a step potential (lower panel; note the different scale in left) show features incompatible with the observed data.  The width of the central region is adjusted to be 46-48~nm in the simulations in order to match the phase of the zero field oscillations.  The amplitude is fit by setting the mean free path in the Landauer formula to be 67~nm in the top panel, 60~nm in the central panel, and 300~nm for in the lower panel.  {\bf c} Transmission probability as a function of angle at zero magnetic field for the pn junctions with (red, solid) and without (green, dotted) nonlinear screening, and for the step potential (black, dashed).}
\label{figs1} \end{figure}


\subsection*{Disordered heterojunctions}
Several measured devices showed Fabry-Perot resonances; data from a second device (the same used in Fig. 3d of the main text), is shown in Fig~6.  Most of features discussed in text are present, including an oscillatory conductance that can be tuned be magnetic field. In this device, however, as in the majority of devices, the Fabry-Perot resonances appear to be mixed with other, irregular, conductance fluctuations.  This behavior is most evident in the magnetic field dependence of the oscillations (Fig. 6c): although the phase shift is still evident in some of the transmission resonances, several oscillation periods appear to be intermixed, and there is no adiabatic crossover to the Shubnikov-de Haas oscillations as observed in the high quality device discussed in the main text.  In addition, we discovered that the more disordered heterojunctions devices exhibit transmission resonances even when the overall doping of the LGR was of the same sign as that in the GL (Fig. 6b).  We interpret these effects as a combination of higher disorder concentration between the p-n junctions causing enhanced universal conductance fluctuations and inhomogeneous gate coupling, which has the effect of causing an averaging over several Fabry Perot fringes.~\cite{levitov_2008}

\newpage
\begin{figure}[h]
\centering
\includegraphics[width=120mm]{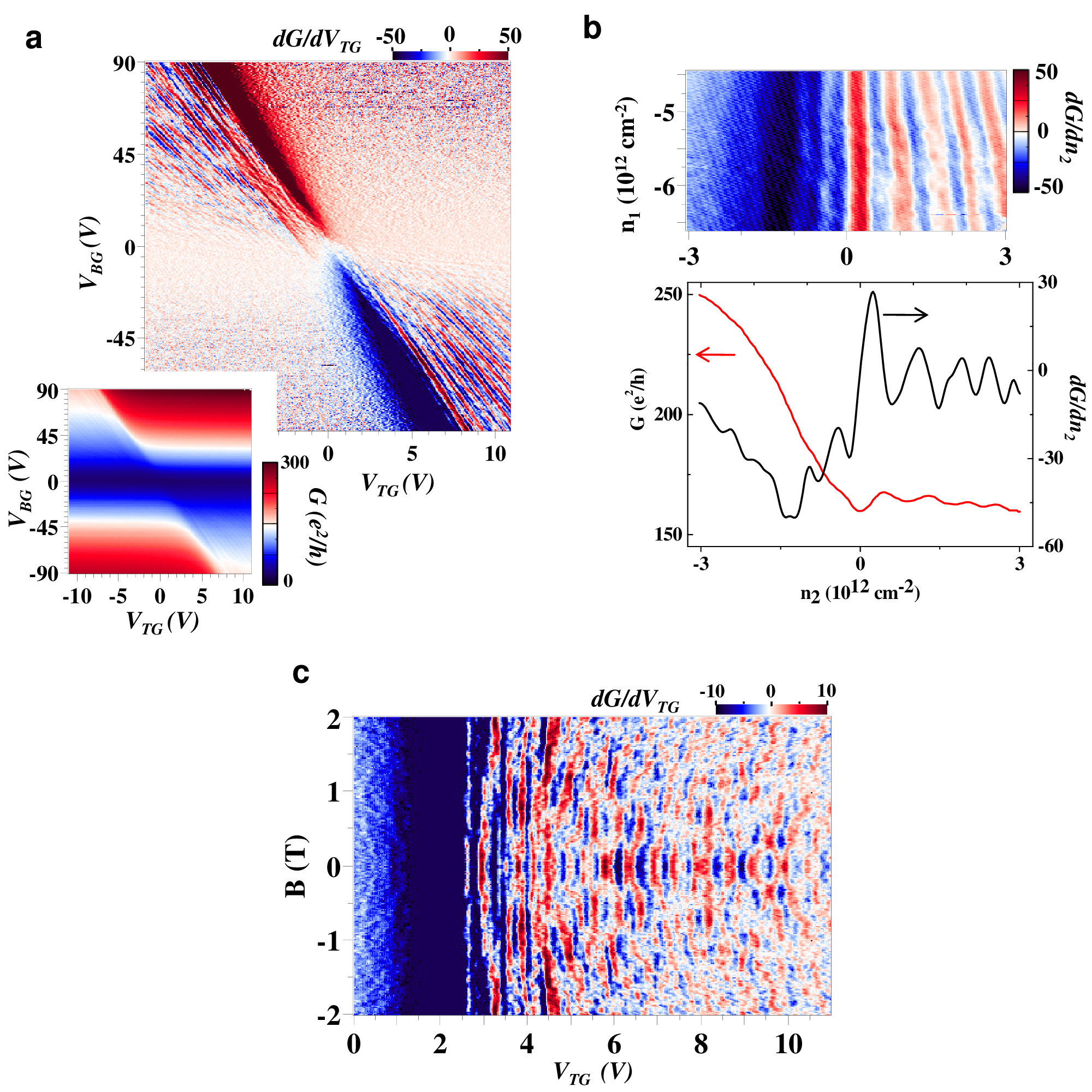}
\caption{Data from a second device.  {\bf a} Inset:  Conductance, measured in the four wire geometry, as a function of top and back gate voltages.  Main Panel:  Derivative of conductance with respect to top gate voltage (arbitrary  units).
{\bf b} Oscillations in the unipolar regime.  The oscillations are visible in the derivative of the conductance with respect to top gate voltage, and as kinks in the steeply falling background conductance as the central region approaches charge neutrality.  {\bf c} The irregular character of the oscillations becomes apparent in the magnetic field dependence of the oscillation extrema.  Although the phase shift is still visible, several periods seem intermixed, and there is no apparent adiabatic connection to the Shubnikov-de Haas oscillations.}
\end{figure}

\end{document}